\documentclass[lettersize,journal]{IEEEtran}
\usepackage{amsmath,amsfonts}
\usepackage{subcaption}
\usepackage{algorithm}
\usepackage{array}
\usepackage{textcomp}
\usepackage{stfloats}
\usepackage{url}
\usepackage{verbatim}
\usepackage{graphicx}
\usepackage{cite}
\usepackage{xcolor}
\newcommand\revision[1]{\textcolor{black}{#1}}

\usepackage{tikz}

\usepackage{algpseudocode}
\usepackage{siunitx}
\usepackage{multirow, makecell}
\usepackage{caption}
\usepackage{accents}
\usepackage{enumitem}

\newcommand{\myeq}{\textbf{== }}

\newcommand{\ignore}[1]{}
\newcommand{\ubar}[1]{\underaccent{\bar}{#1}}

\begin{document}
\title{Tracking Real-time Anomalies in Cyber-Physical Systems Through Dynamic Behavioral Analysis}

\author{Prashanth Krishnamurthy,
\and
Ali Rasteh,
\and
Ramesh Karri,
\and
Farshad Khorrami
\thanks{This work was supported in part by DOE NETL (DE-CR0000017) and NSF SaTC
  (2039615). The authors are with the Control/Robotics Research Laboratory (CRRL), Dept. of Electrical and Computer Engineering, NYU Tandon School of Engineering, Brooklyn, NY 11201, USA.}
}

\maketitle

\begin{abstract}
    Increased connectivity and remote reprogrammability/reconfigurability features of embedded devices in current-day power systems (including interconnections between information technology -- IT -- and operational technology -- OT -- networks) enable greater agility, reduced operator workload, and enhanced power system performance and capabilities. However, these features also expose a wider cyber-attack surface, underscoring need for robust real-time monitoring and anomaly detection in power systems, and more generally in Cyber-Physical Systems (CPS). The increasingly complex, diverse, and potentially untrustworthy software and hardware supply chains also make need for robust security tools more stringent. We propose a novel framework for real-time monitoring and anomaly detection in CPS, specifically smart grid substations and SCADA systems. The proposed method enables real-time signal temporal logic condition-based anomaly monitoring by processing raw captured packets from the communication network through a hierarchical semantic extraction and tag processing pipeline into time series of semantic events and observations, that are then evaluated against expected temporal properties to detect and localize anomalies.  We demonstrate efficacy of our methodology on a hardware in the loop testbed, including multiple physical power equipment (real-time automation controllers and relays) and simulated devices (Phasor Measurement Units -- PMUs, relays, Phasor Data Concentrators -- PDCs), interfaced to a dynamic power system simulator. The performance and accuracy of the proposed system is evaluated on multiple attack scenarios on our testbed.
\end{abstract}

\section{Introduction} \label{sec:Introduction}
The smart grid is the next generation of power systems offering promises of a wide variety of benefits in efficiency, reliability, and safety. Some of the key features of smart grids are agile reconfigurability and dynamic optimization of grid operations, rapid detection and response to faults in the system, integration of renewable power sources with conventional fossil fuels, and providing of pervasive monitoring facilities for power systems.
An important step in Industrial Revolution 4.0 is the digitization of Industry 3.0 and bringing together the Information and Communication Technology (ICT) and Operational Technology (OT) for controlling the physical processes, their monitoring, and maintenance \cite{faheem2018smart, tantawi2019advances, nafees2023smart}. In the case of power systems, smart grids are the emerging point of Industrial Revolution 4.0. IT systems in the smart grid technology include different ICT servers, communication technology, supervisory and monitoring infrastructure, etc. On the other hand, OT comprises Programmable Logic Controllers (PLCs), Remote Terminal Units (RTUs), Intelligent Electronic Devices (IEDs), Phasor Measurement Units (PMUs), relays, Human Machine Interfaces (HMIs), etc. \cite{nafees2023smart}. Seamlessly combining ICT and OT can provide efficient methods to augment the capabilities of smart grids.
However, the resulting expanded connectivity and remote programmability/reconfigurability can broaden the attack surface and increase cybersecurity vulnerabilities. Recent examples of attacks on power grids and industrial systems show the crucial importance of these systems and extensive impacts that can result if the security of these systems is compromised. For example, the coordinated cyber-attack on the Ukrainian power grid in 2015 caused a power loss for 6 hours, affecting about 225,000 customers in Ukraine \cite{alert2016cyber}. As another example, the well-known Stuxnet malware was used to attack nuclear industrial systems in 2010, which is known to use zero-day covert attacks \cite{singer2015stuxnet}. Thus, ensuring the security of these systems is of paramount importance.

\begin{figure}[!t]
  \centering
  \includegraphics[width=0.87\linewidth]{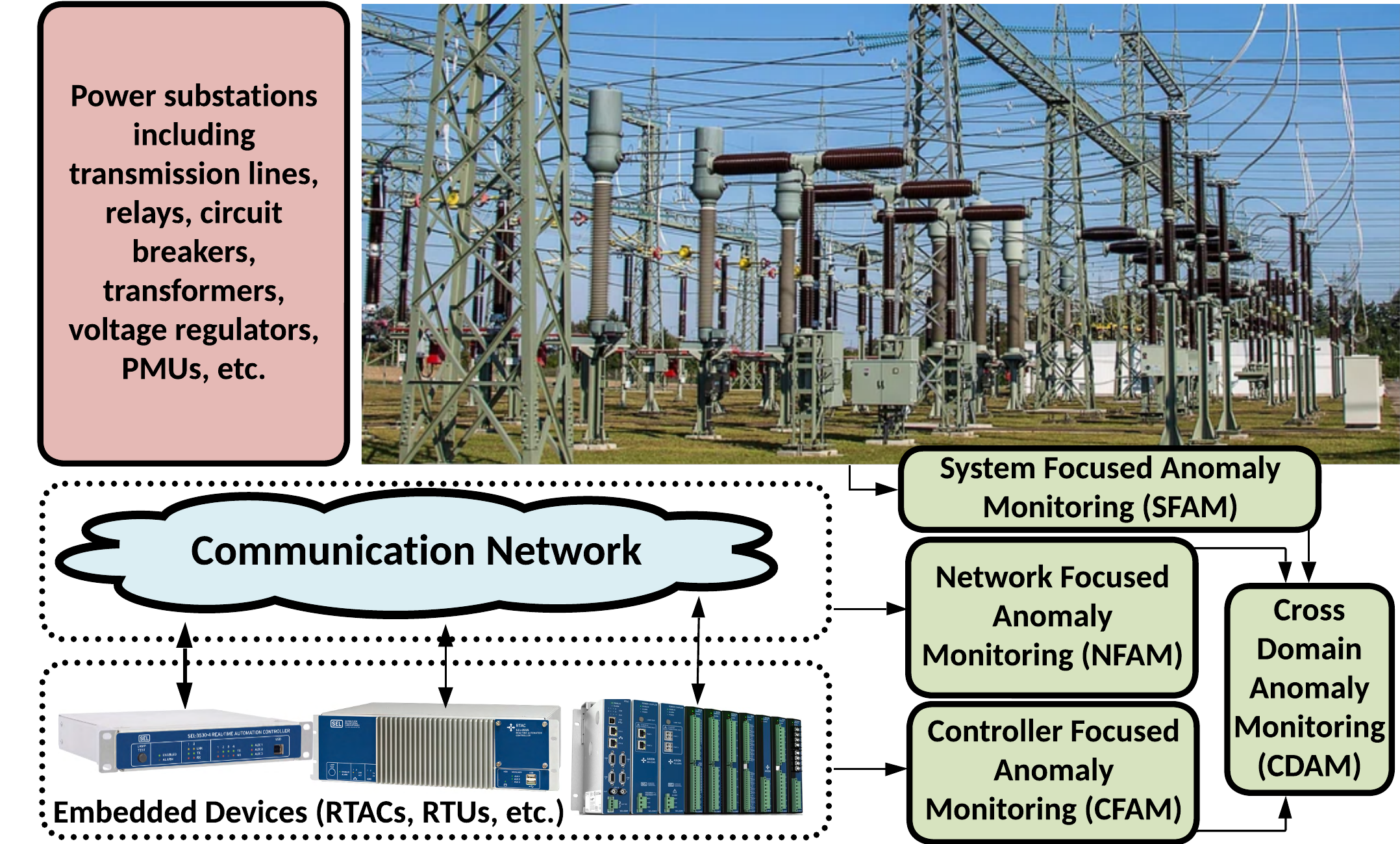}
  \caption{Proposed TRAPS multi-domain monitoring approach with a unified framework for network-focused anomaly monitoring (NFAM), controller-focused anomaly monitoring (CFAM), system-focused anomaly monitoring (SFAM), and cross-domain anomaly monitoring (CDAM).  
  }
  \label{fig:grid_cfam_nfam_sfam_cdam}
\end{figure}

Industrial control systems, including smart grids, are complex systems consisting of embedded device nodes interconnected by communication networks and interfaced to physical processes. Relevant devices for smart grids include, for example, MTUs (Master Terminal Units), RTUs, RTACs (Real-Time Automation Controllers), PLCs, relays, PMUs, PDCs (Phasor Data Concentrators), and HMI. These devices communicate using various protocols such as DNP3, IEEE C37.118, Modbus, IEC61850, SEL Fast Msg, OPC-UA (Open Platform Communications Unified Architecture), IEC 60870-5, etc., which are all industrial communication standards primarily developed several decades back without specific focus on security. The temporal evolution of the cyber-physical system is governed by the device behaviors (e.g., logic/rules programmed on the devices), the communications/interactions between the devices, and the physical dynamics of the system. A malicious manipulation of a device behavior or of the communication network (e.g., device spoofing, packet injection or manipulation, etc.) by an intruder/adversary can lead to catastrophic consequences in the power grid, including destabilization of the grid and damaging physical components in the grid. 
Therefore, techniques to monitor these interactions and processes in real time and flag any anomalies with low latency and high accuracy would be vitally beneficial to security of the grid. Such a technology should not only consider the basic communication specifications between the grid devices but also whether the observed temporal processes are consistent with the expected behaviors and dynamics of the grid. Since the smart grid is a composite of the controller devices, power system dynamics, network communication channels, and interplay between these components, a comprehensive monitoring system should be able to track the temporal behaviors in real time and detect any abnormalities. In analogy with anomaly monitoring systems for other cyber-physical systems \cite{khorrami2016cybersecurity} and autonomous vehicles \cite{patel2020learning}, such a comprehensive monitoring system should span 
controller-focused anomaly monitoring (CFAM) for validating behaviors of controllers and other devices in the grid, network-focused anomaly monitoring (NFAM) for validating network-level transactions and statistics, system-focused anomaly monitoring (SFAM) for validating temporal process dynamics, and cross-domain anomaly monitoring (CDAM) for validating interplay between  controller/system/network components (Figure \ref{fig:grid_cfam_nfam_sfam_cdam}).

To address this crucial need, we develop a real-time integrity verification
methodology (TRAPS -- \underline{T}racking \underline{R}eal-time
\underline{A}nomalies in cyber-\underline{P}hysical \underline{S}ystems) in this
paper to detect abnormal behaviors in the power system by continuous dynamic
behavioral analysis of the cyber-physical system. The proposed TRAPS approach is
based on a signal temporal logic (STL) condition-based anomaly monitoring and
Intrusion Detection System that processes real-time observations from
communication network packet captures through a hierarchical semantic extraction
and tag processing pipeline to transform them into a time series of semantic
events and observations, collectively referred to as semantic tags. These
semantic tag time series are then evaluated against expected temporal properties
to detect and localize anomalies and visualize them in a dashboard graphical
user interface (GUI).  We demonstrate the efficacy of our proposed methodology with several attack scenarios on a hardware-in-the-loop (HIL) testbed, which includes both physical and virtual power devices, interfaced to a dynamic power system simulator.

\section{Related Works} \label{sec:Related_Works}
Several types of attacks on smart grid systems have been considered in the
literature (e.g., \cite{wlazlo2021man}) including Measurement Integrity Attacks, False Data Injection (FDI),
False Command Injection (FCI), Control Logic Modification, and Denial of Service (DoS)
Attacks including Time-Delay and Jamming Attacks. Coordinated attacks, which use
multi-stage complicated patterns to increase the effectiveness, and Cascading
attacks, which use a Single Point of Failure to propagate the effect to the
other points of the system have also been considered \cite{nafees2023smart}.
To defend against the various attacks, defense approaches (termed
in general as Intrusion Detection Systems or Anomaly Detection Systems) have
been developed as discussed below.

\noindent{\bf Signature-Based Methods} use a ``blacklist'' of signatures of
prior attack/anomaly events to detect intrusions of the same category, but
cannot detect unknown/zero-day attacks with new signatures. Examples of 
methods of this type include \cite{wlazlo2021man} which detected
machine-in-the-middle (MITM) attacks on the DNP3
protocol through Snort rules, \cite{sahu2021multi} which applied ML-based fusion
of cyber and physical sensors to detect FCI/FDI attacks on DNP3 protocols, and
\cite{kang2016towards} which applied Suricata, an open-source network IDS, to
detect anomalies for network protocols, including IEC61850, based on software
rules.

\noindent{\bf Specification-Based Methods} model the system's behavior using its
specifications, especially at the network level, and analyze the observed
behavior such as the communication protocol details to detect abnormalities. %
Typical limitations in the available methods of this type include support for only specific protocols, limited scalability, and ability to monitor only specific types of behaviors/events.
Specification/behavior based IDS have been developed considering various
protocols such as IEEE C37.118 \cite{sprabery2013protocol,pan2015specification},
DNP3 \cite{lin2013adapting}, IEC60870 \cite{yang2013intrusion}, IEC61850
\cite{premaratne2010intrusion,kwon2015behavior}, and Modbus/TCP \cite{cheung2007using}.
Combinations of multiple IDS approaches have also been studied such as: combination of signature-based and model-based methods using Snort in \cite{yang2013intrusion}; combination of access-control, protocol whitelisting, model-based, and multi-parameter-based detection methods in \cite{yang2016multidimensional}; combinations of host-based and network-based detection methods in \cite{hong2014integrated}; combination of access control, protocol-based, and behavioral whitelists \cite{yang2014multiattribute}. Other specification-based approaches in the literature include monitoring of values of process variables in terms of rules defined in a specific description language in \cite{nivethan2016scada}, state tracking methods \cite{fovino2010modbus,carcano2011multidimensional}, and sequence of events monitoring using a Discrete-Time Markov Chain model in \cite{caselli2015sequence}. Moving-target defense methods in the context of state estimators have also been studied \cite{liu2021optimal, tian2019moving}.

\noindent{\bf Learning-Based Methods} use data-driven machine learning
to detect anomalous or abnormal patterns in the system's traffic/signals.
Challenges when applying these methods include difficulty in obtaining extensive
training datasets, lack of explainability of ML prediction results that can also make it difficult to localize underlying causes of detected anomalies and guide appropriate remediations, and limitations in generalizability under changes in data distribution (domain shift).
Learning-based methods have been applied, for example, to detection of false data injection attacks \cite{he2017real,zhang2020detecting} and time delay attacks \cite{ganesh2021learning} and anomaly detection in transmission protective relays \cite{khaw2020deep}, wide-area protection systems \cite{singh2021cyber}, and distribution systems \cite{cui2020flexible}. Host-based anomaly detectors using analog/digital side channels such as Hardware Performance Counters (HPCs) have been developed (e.g., \cite{krishnamurthy2020anomaly,konstantinou2022hpc}).

In contrast to the various approaches outlined above, the key benefits of the proposed TRAPS approach are:
a unified framework for processing of at-scale heterogeneous communication traffic for monitoring of an open and extensible set of behavioral properties defined through the hierarchical semantic tag processing and STL-based monitoring approach,
end-to-end structure ranging from the ingest of raw network packet captures to semantic parsing, situational awareness, and anomaly detection with graphical visualization,
and computational simplicity and scalability enabling real-time processing of high-bandwidth traffic and simultaneous monitoring of several hundreds of tags and STL conditions.

\begin{figure}[!t]
  \centering
  \includegraphics[width=0.82\linewidth]{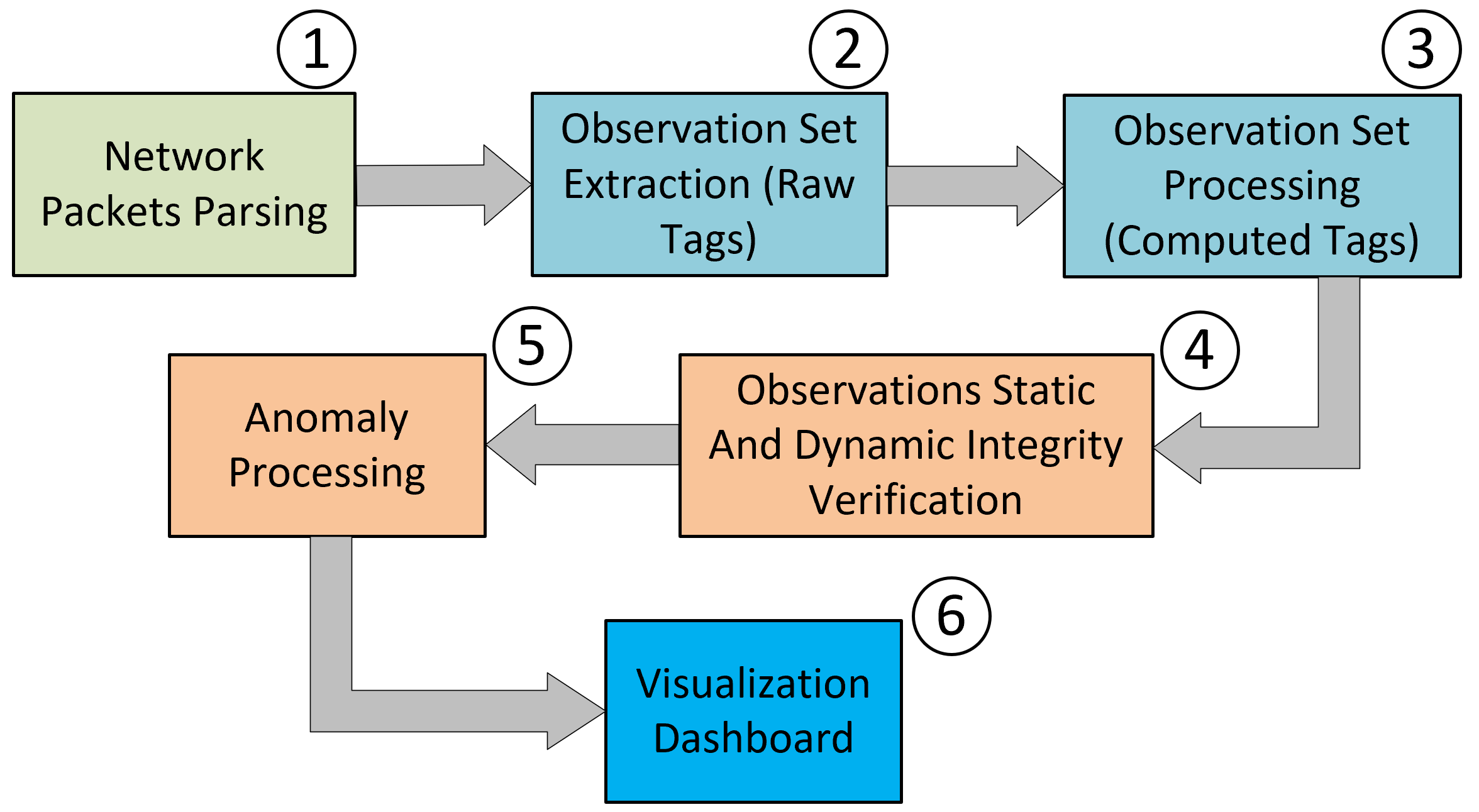}
  \caption{Overall structure of the proposed method. 
  }
  \label{fig:General_block_diagram}
\end{figure}

\section{The Proposed Method} \label{sec:The_Proposed_Method}
The proposed framework is based on the pipeline architecture shown in Figure
\ref{fig:General_block_diagram}, where raw data is processed through a
sequence of layers to extract time series observations of hierarchically defined
semantic tags, that are then used for anomaly detection relative to a set of
STL-based behavioral specifications, and visualization in an operator dashboard.
The individual components are discussed in the subsections below.

\subsection{Network Packets Parsing} \label{sec:Network_Packets_Parsing}
The raw network traffic (either live or as a pcap) that is the input to TRAPS comprises of the communications between the various devices in the smart grid using several different protocols such as the following supported by our current implementation of our system: DNP3, IEEE C37.118, Modbus, IEC61850 GOOSE, IEC61850 MMS, and SEL Fast Msg \revision{(a proprietary protocol by Schweitzer Engineering Laboratories Company)}, OPC-UA, and Telnet protocols. A set of scripts based on open-source libraries such as Scapy and Pyshark are used to process the network traffic to parse and extract the payload contents using methodologies analogous to \cite{anantharaman2018phasorsec,anantharaman2022communications}.
For efficient real-time processing, the implementation is structured as a set of separate Docker containers for each communication protocol. A front-end ingest module detects the application layer protocol for each incoming packet and forwards it to the appropriate protocol-specific parser for extracting the payload contents.
The outputs of the protocol-specific parsers are combined into an MQTT streaming feed that is then used by the semantic tag processing component. In addition to the MQTT feed, the combined output stream from the parsers is exposed via a REST API interface.
Both the push (streaming) and pull (REST API) interfaces to the parser outputs support filtering on properties such as IP addresses, protocols, and message types.
\subsection{Observation Set Extraction} \label{sec:Observation_Set_Extraction}
The output of the protocol-specific parsers is a time series ${\cal P}$ of records of form $P_i = (t_i, s_i, d_i, p_i, m_i), i\in[1,n]$ where:
\begin{itemize}
\item $t_i$ is the packet's timestamp
\item $s_i$ and $d_i$ are the source and destination of the packet, respectively, which could be IP and/or MAC addresses depending on the protocol
\item $p_i$ is the protocol and message type of the packet
\item $m_i$ is the set of measurements/values in the packet's payload such as analog and digital values in IEEE C37.118, Modbus coils, Modbus holding registers, DNP3 analog inputs and outputs, etc.
\end{itemize}
The specific information mapping (i.e., which fields in a DNP3 message correspond to what physical quantities) are installation-specific and can vary widely.
Hence, after parsing of raw fields in the network packets, a key step is
mapping of the fields to semantic variables. For this purpose, TRAPS uses a
flexible query set structure wherein functions defined over the raw fields are
used to populate the values of semantic variables as appropriate for the
particular installation and the particular network communication protocol. For
example,  
in IEEE C37.118, the constituent fields are typically phasors, analogs (e.g.,
currents and voltages), and digitals (e.g., status values).
The queries to extract semantic variables (``raw tags'') from the time series ${\cal P}$ is defined as a set of packet filtering rules of form $R_i = (s_i, d_i, p_i, a_i, st_i), i\in[1,m]$ where $s_i$, $d_i$, $p_i$ have the same meaning as in ${\cal P}$, $a_i$ is an attribute address specifier (e.g., index of the data in DNP3 binary inputs/outputs, address of an input register in Modbus, etc.), and $st_i$ is a tag identifier to be raised whenever the filtering rule is triggered due to a matching $(s_i,d_i,p_i,a_i)$.
The algorithmic structure of this component is shown in Algorithm~\ref{alg:raw_tags}.

\begin{algorithm}
\caption{Filtering time series of parsed packets to generate time series of raw tags.}\label{alg:raw_tags}
\begin{algorithmic}[1]
\For{packet $P_i$ in parsed packets}
    \For{j = 1 to m}
        \State $R_j = (s_j, d_j, p_j, a_j, st_j)$
        \If{$(s_i,d_i,p_i) \myeq (s_j,d_j,p_j)$}
            \If{attributes addressed by $a_j$ exist in $P_i$}
                \State Extract attributes addressed by $a_j$ from $P_i$
                \State Push $st_j$ into output time series
            \EndIf
        \EndIf
    \EndFor
\EndFor
\end{algorithmic}
\end{algorithm}

\subsection{Observation Set Processing} \label{sec:Observation_Set_Processing}
Since the behavioral properties of the CPS might be most naturally described not
in terms of raw tags but in
terms of variables that are computed as functions of multiple tags over multiple
time instants. Hence, TRAPS includes a hierarchical tag processing engine that
allows definitions of computed tags as functions of other raw/computed tags. To facilitate a flexible structure for defining hierarchical dependencies of computed tags, a Directed Acyclic Graph (DAG) ${\cal D}$ is used in which each node represents a time series of a particular tag and is constructed as a functional of the previously extracted time series of tags in the node's dependency list. The functional dependency structure is represented as a set of filtering rules of form
$C_k=(Dep_k, f_k, st_k),k\in[1,s]$ where
$Dep_k$ denotes the dependency list (of raw/computed tags), $f_k$ is a function encoding the calculations required to obtain $C_k$ from the time series values of the tags in the dependency list, and $st_k$ is a tag identifier to be raised whenever the filtering rule is triggered, similar to the corresponding designator for raw tags.
The algorithmic structure of this component is shown in Algorithm~\ref{alg:computed_tag_list} where the input queue ${\cal Q}$ holds both raw tags raised from Algorithm~\ref{alg:raw_tags} and computed tags pushed as part of Algorithm~\ref{alg:computed_tag_list} (to iteratively process downstream dependencies in the DAG).
The time series of observations for each semantically extracted tag is of form $P =
\{(t_i,v_i)\}_{i=1,\ldots,n}$ with $t_i, v_i$ being the timestamp and value,
respectively, of that tag.

\begin{algorithm}
\caption{Extracting time series of computed tags.}\label{alg:computed_tag_list}
\begin{algorithmic}[1]
  \While{True}
  \State Get next tag $q$ from queue ${\cal Q}$ (or wait until there is one).
  \For{each child node $k$ of $q$ in DAG ${\cal D}$}
  \State Compute $f_k$ using time series values of tags from $Dep_k$ and add computed value to time series of observations for tag $st_k$.
  \State Push $st_k$ to ${\cal Q}$.
  \EndFor
\EndWhile
\end{algorithmic}
\end{algorithm}

\subsection{Observation Set Static \& Temporal Integrity Verification} \label{sec:Tags_Verification}
A crucial property of the CPS is that its expected behavior (as defined by
device logic, system dynamics, etc.) implies various correlations/causalities among the
time series of tags.
These include both dependencies/relationships between values of two time series
(e.g., expected relationships between relay open/closed status and voltage
values) and temporal properties (e.g., an event in one time series expected to
happen before or after an event in a second time series).
These relations could stem from physics-based and behavior-based properties of
the CPS. For example, physics-based properties result from power system physical
laws such as the Kirchhoff's voltage and current laws (e.g., interdependencies
between PMU measurements at different locations) 
while behavior-based properties relate to device configurations, network
characteristics, etc. For example, in the case of devices such as proxies, protocol converters,
 or command forwarders in the smart grid, the values of corresponding time series from 
 before-proxy and after-proxy traffic have expected relationships, both in terms
 of numerical values of payload contents and  temporal relationships
 between messages (e.g., time delays before retransmission).
 Deviations from expected correlation/causality relations
 indicate anomalies or abnormal behavior that could stem from cyberattacks,
 physical attacks, or physical malfunctions.
 
 To enable flexible monitoring spanning these various types
 of correlation/causality relations, we define several types of condition
 structures discussed below that could hold between time series of different tags. To show
 examples of the
 condition structures, we use the notations
 $P_1 = \{(t_{1,i},v_{1,i})\}_{i=1,\ldots,n}$, $P_2 =
 \{(t_{2,j},v_{2,j})\}_{j=1,\ldots,m},\ldots$,
 $P_r=\{(t_{r,k},v_{r,k})\}_{k=1,\ldots,q}$  to denote the time series of
 observations of various tags.
 \begin{itemize}
   \item {\em threshold conditions} such as
\begin{align}
  |v_{1,i}-v_{1,i-1}| &\leq V_{th}
  \label{eq:threshold_value}
\\
\ubar T_{th} &\leq|t_{1,i}-t_{1,i-1}|\leq \bar T_{th}
\label{eq:threshold_time}
\end{align}
where
$V_{th}$, $\ubar T_{th}$, and $\bar T_{th}$ denote the value threshold, 
lower timing threshold, and upper timing threshold, respectively, for the observations.
\item {\em match conditions} such as
\begin{equation}\label{eq:match}
\begin{split}
|v_{1,i'}-v_{2,j'}|\leq V_{th}
\end{split}
\end{equation}
where 
$i'$ and $j'$ denote matching time instants of the time series of observations $P_1$ and $P_2$ (e.g., time values such that
$t_{1,i'}\approx t_{2,j'}$). More generally, functional match conditions (with an
arbitrary function $f$) across
multiple time series of observations can be defined of the form
\begin{equation}
    f(v_{1,i'}, v_{2,j'}, \cdots v_{r,k'}) =0
  \label{eq:functions}
\end{equation}
where
$i',j',\ldots,k'$ denote matching time instants across the different time series
of observations (e.g., time values such that
$t_{1,i'}\approx t_{2,j'}\approx \ldots \approx t_{r,k'}$).
\item {\em pre-conditions}, i.e., conditions that an event (defined in general in
  terms of
  values from one or more time series of observations) should have been preceded
  by some other defined event within some time interval; for example,
  \begin{align}
    f_1(v_{1,i})=0 \Longrightarrow \exists &t_{2,j} \in [t_{1,i}-T_{th},t_{1,i})
                                             \nonumber\\
    & s.t. f_2(v_{1,i},v_{2,j}) = 0
\label{eq:precondition}
\end{align}
where $f_1$ and $f_2$ are arbitrary functions and $T_{th}$ is a threshold
on timing; for example, a condition that the time series of observations $P_1$
should track (possibly with a delay) the time series of observations $P_2$ would
be represented with $f_1(v_{1,i})\equiv 0$ and
$f_2(v_{1,i},v_{2,j})=\max\{|v_{1,i}-v_{2,j}|-V_{th},0\}$ where $V_{th}$ is a
threshold for the matching of $v_{1,i}$ and $v_{2,j}$.
\item {\em post-conditions}, i.e., conditions that some specified event should be followed
  by some other defined event within some time interval; for example,
  \begin{align}
    f_1(v_{1,i})=0 \Longrightarrow \exists &t_{2,j} \in [t_{1,i},t_{1,i}+T_{th})
                                             \nonumber\\
    &s.t. f_2(v_{1,i},v_{2,j}) = 0
  \label{eq:postcondition}
\end{align}
where $f_1$ and $f_2$ are arbitrary functions and $T_{th}$ is a threshold
on timing.
\end{itemize}

The algorithmic structure of the integrity verification component for flagging condition violations is shown in Algorithm~\ref{alg:condition_check} where ${\cal C}$ denotes the set of all conditions defined in a particular deployment configuration.
  Besides the example structures above, the conditions can involve
  dependencies on arbitrary numbers of tags as well as time
  history of the tags. Also, the conditions can involve other similarity
  measures between time series such as $L_p$ norms/distances, dynamic time
  warping, correlation measures, etc. 

\begin{algorithm}
\caption{Algorithm for flagging condition violations.}\label{alg:condition_check}
\begin{algorithmic}[1]
  \For{condition $c\in{\cal C}$}
  \If{deviation from condition check as in \eqref{eq:threshold_value}-\eqref{eq:postcondition}}
  \State Push anomaly detection flag on $c$ to anomaly queue ${\cal M}$ with metadata on timestamp and variables used in computation of condition $c$
  \EndIf
  \EndFor
\end{algorithmic}
\end{algorithm}

\subsection{Anomaly Localization} \label{sec:Anomaly_localization}
Each raw tag and computed tag maintains a provenance information as to which specific underlying communication observation was involved in the observed value of the particular tag. Hence, considering the devices in the CPS and the observed communications between
devices as a communication graph ${\cal G}$, flagging of a condition violation directly indicates
potential physical locations of the anomaly based on the edges (communication
links) related to the constituent underlying tags in the condition check and the
corresponding adjoining nodes.
The DAG structure of raw and computed tags enables efficient retrieval of
underlying raw tags corresponding to any flagged anomaly.
Hence, anomaly scores are maintained for each
node and edge and these scores are incremented each time a related anomaly is
flagged.
To enable the operator to rapidly see the most likely anomalous nodes/edges, the
anomaly scores are normalized over the graph ${\cal G}$ and used for color
coding in a graphical visualization (Figure
\ref{fig:Grafana_nodes_graph}). The algorithmic structure of this component is
shown in Algorithm~\ref{alg:anomaly_score}.

\begin{algorithm}
\caption{Algorithm for anomaly provenance scoring.}\label{alg:anomaly_score}
\begin{algorithmic}[1]
\State Set anomaly scores to 0 $\forall$ nodes and edges in graph ${\cal G}$.
\For{$M_i$ in anomaly queue ${\cal M}$}
    \State Look up all underlying raw tags $R_j$ for $M_i$ using DAG.
    \For{$R_j$ in effective raw tags}
        \State Look up corresponding nodes and edge for $R_j$ and increment their
        anomaly scores.
    \EndFor
\EndFor
        \State Normalize anomaly scores for nodes and edges so that $\sum_{n\in
          {\cal N}}a(n)=1$ and $\sum_{e\in
          {\cal E}}a(e)=1$ where ${\cal N}$ and ${\cal E}$ are the sets of nodes
        and edges in graph ${\cal G}$ and $a(.)$ is the anomaly score for a node/edge.
\end{algorithmic}
\end{algorithm}

\subsection{Visualization Dashboard} \label{sec:Visualization_Dashboard}
\begin{figure}[!t]
  \centering
  \includegraphics[width=1.0\linewidth]{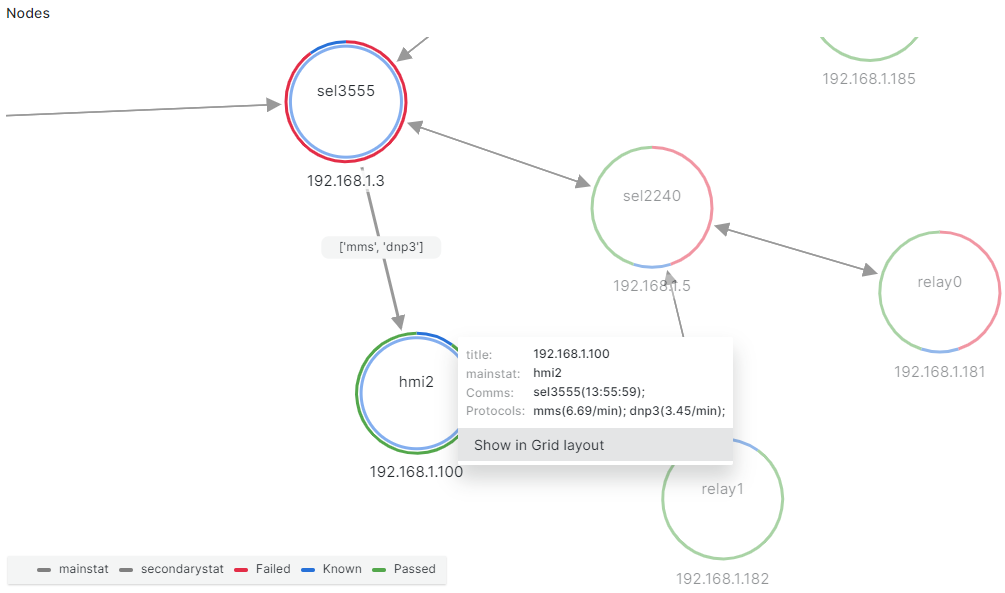}
  \caption{Sample screenshot of automatically generated nodes and edges graph in our Grafana-based GUI dashboard.}
  \label{fig:Grafana_nodes_graph}
\end{figure}

The user front-end of TRAPS is a dashboard GUI implemented using Grafana to
visualize summaries of detected anomalies and overall semantically
parsed observations with a hierarchical interactive interface providing an
easy-to-use top-level summary as a broad overview and on-demand interactive
mechanisms to access additional details when desired.
The dashboard shows various elements of situational awareness and anomaly
detection, such as observed nodes and communications (along with salient communication properties such as request/response timing), tag values and tag histories, and detection of anomalies in expected match/pre/post conditions and the provenance of detected anomalies.
Also, a graph of the network architecture with color-coded visualization of
detected anomalies is embedded in the GUI (sample screenshot in Figure
\ref{fig:Grafana_nodes_graph}).
The dashboard also provides plots of tag
histories and tabular views of communications and tag values (screenshots
omitted for brevity).

\section{Experimental Results} \label{sec:Experimental_Results}
\subsection{Experimental Setup And Behavioral Modeling} \label{sec:Setup}
\begin{figure}[!t]
  \centering
  \includegraphics[width=0.55\linewidth]{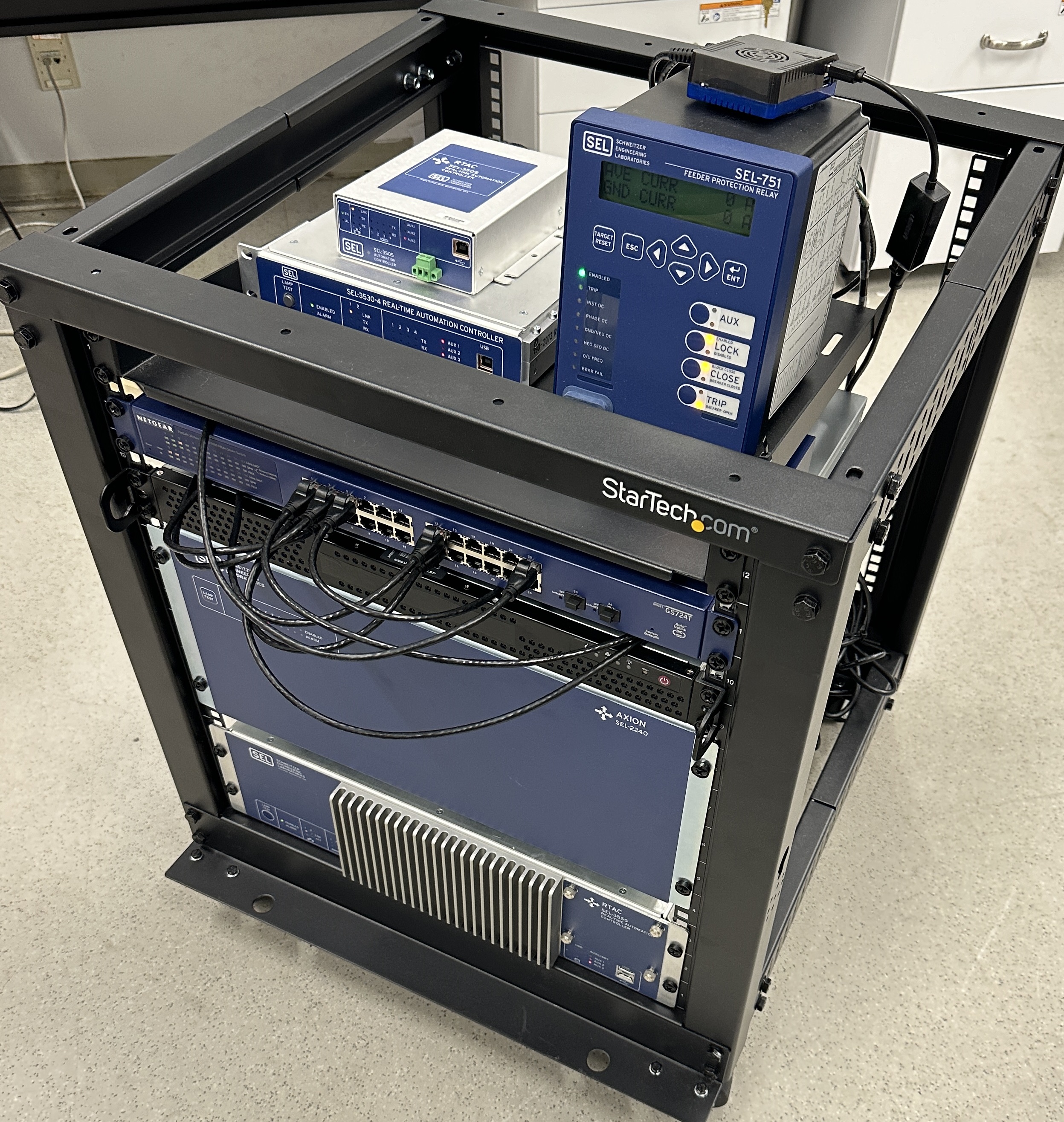}
  \caption{Our HIL experimental testbed.}
  \label{fig:HIL_physical_image}
\end{figure}

\begin{figure}[!t]
  \centering
  \includegraphics[width=0.84\linewidth]{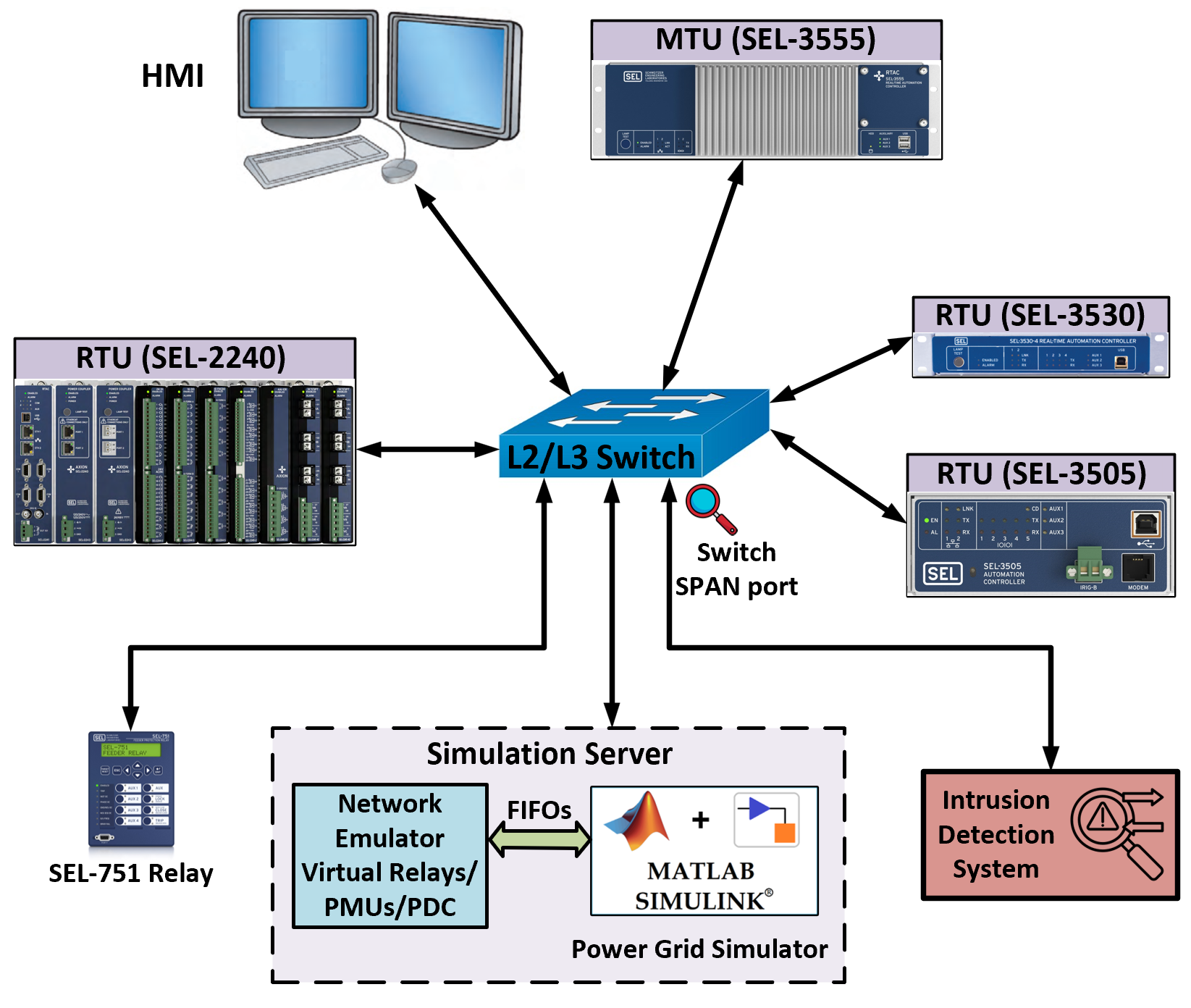}
  \caption{Architecture of HIL testbed.}
  \label{fig:HIL_Physical_setup}
\end{figure}

\begin{figure}[!t]
  \centering
  \includegraphics[width=1.0\linewidth]{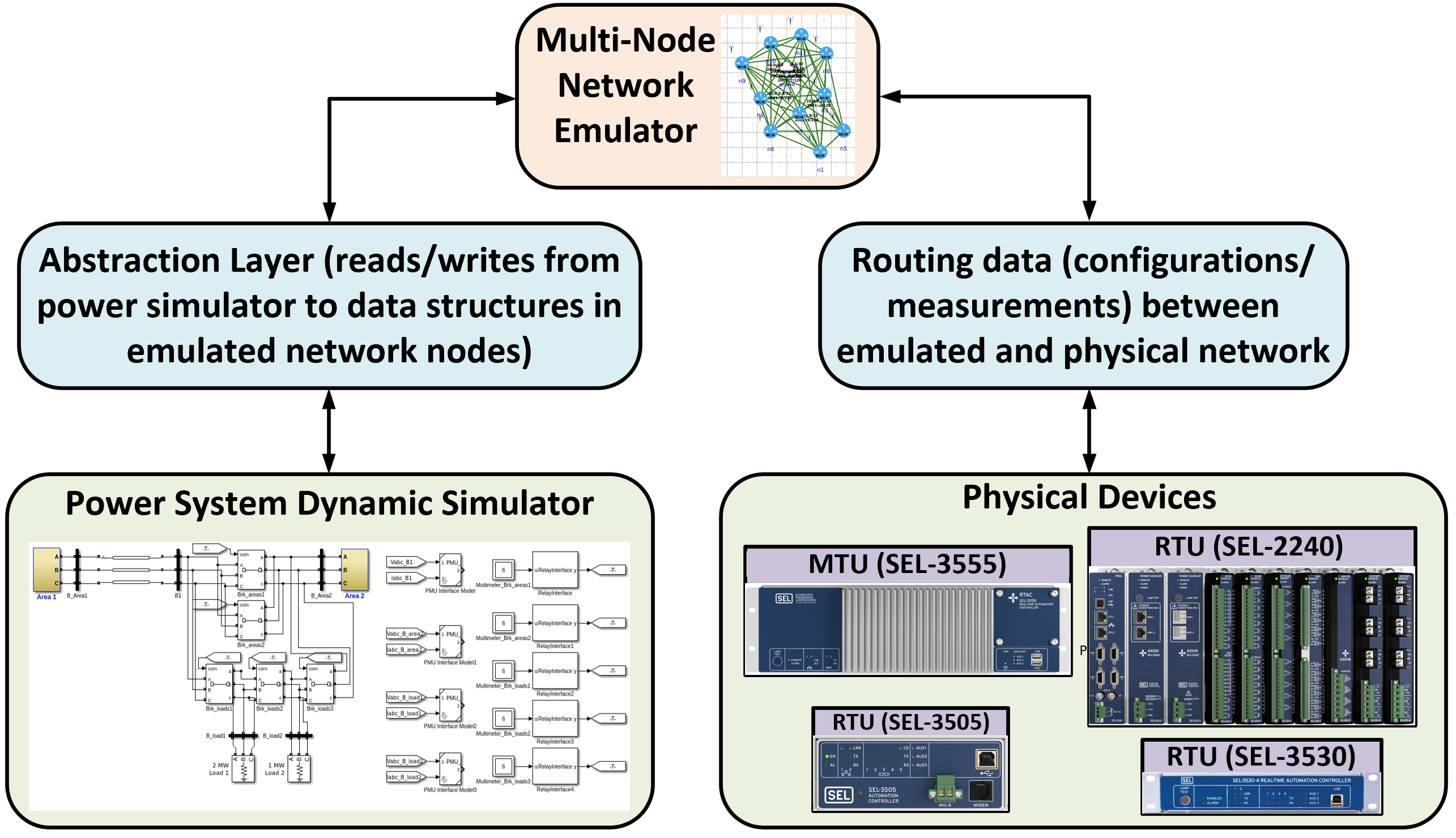}
  \caption{Architecture of communications between physical devices (SEL), emulated devices, and the power system dynamic simulator. The power system simulator uses an abstraction layer to interact with emulated nodes (including Relays, PMUs, and PDC). On the other hand, emulated nodes interact directly with physical nodes in an L2/L3 network.}
  \label{fig:HIL_Physical_Virtual_relation}
\end{figure}

To evaluate the efficacy of the proposed framework, we developed a HIL testbed (Figure~\ref{fig:HIL_physical_image}) with the architecture shown in Figure~\ref{fig:HIL_Physical_setup}.
The power grid simulator implemented using
Matlab and Simulink running on a Linux server
is interfaced with a network emulator based on the open-source CORE (Common Open Research Emulator) tool for building virtual networks. 
The virtual network emulator transparently routes traffic between both physical nodes and virtual nodes in the testbed as illustrated in Figure~\ref{fig:HIL_Physical_Virtual_relation}.
The physical devices in the testbed are SEL (Schweitzer Engineering Laboratories) RTACs, which have been configured to have different roles including a
Human-Machine Interface (HMI), data concentrator, and relay control logic devices.
CORE allows creation of virtual nodes, each of which runs in a separate Linux namespace. Using this functionality, 
virtual devices were defined to simulate virtual IEDs like Relays, PMUs, and PDC, each with a corresponding script running in their Linux namespace to defined their behavior and role.
The MATLAB-based simulator and the network emulator communicate using FIFO (first-in-first-out) special files, via which the status of the relays and PMUs measurements are transferred.
The HMI  designed on the SEL-3555 can be accessed through the web interface and includes graphical functionalities to designed to monitor values, control relays, and trigger attacks in the system for testing TRAPS. All physical devices and the simulation computer are connected using an L2/L3 Netgear switch. The switch's SPAN (Switched Port Analyzer) port is used to mirror all the traffic data in the network and send a copy to the monitoring machine (a Linux workstation), which then analyzes the traffic using the TRAPS  framework described in this paper to detect anomalies. 

\begin{figure}[!h]
  \centering
        \includegraphics[width=0.97\columnwidth]{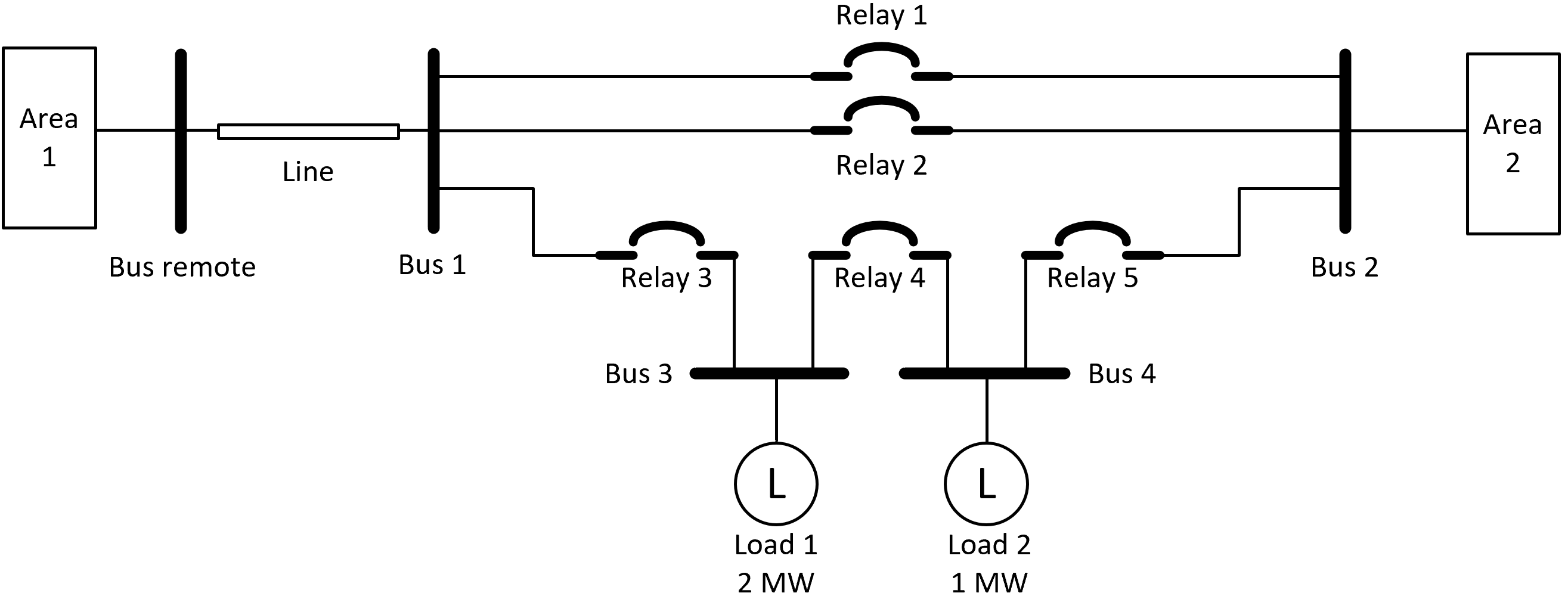}
        \caption{Simple power system topology defined for HIL testing.
        }
  \label{fig:MATLAB_power_system_simulation}
\end{figure}

\begin{figure*}[!t]
  \centering
  \includegraphics[width=0.68\textwidth]{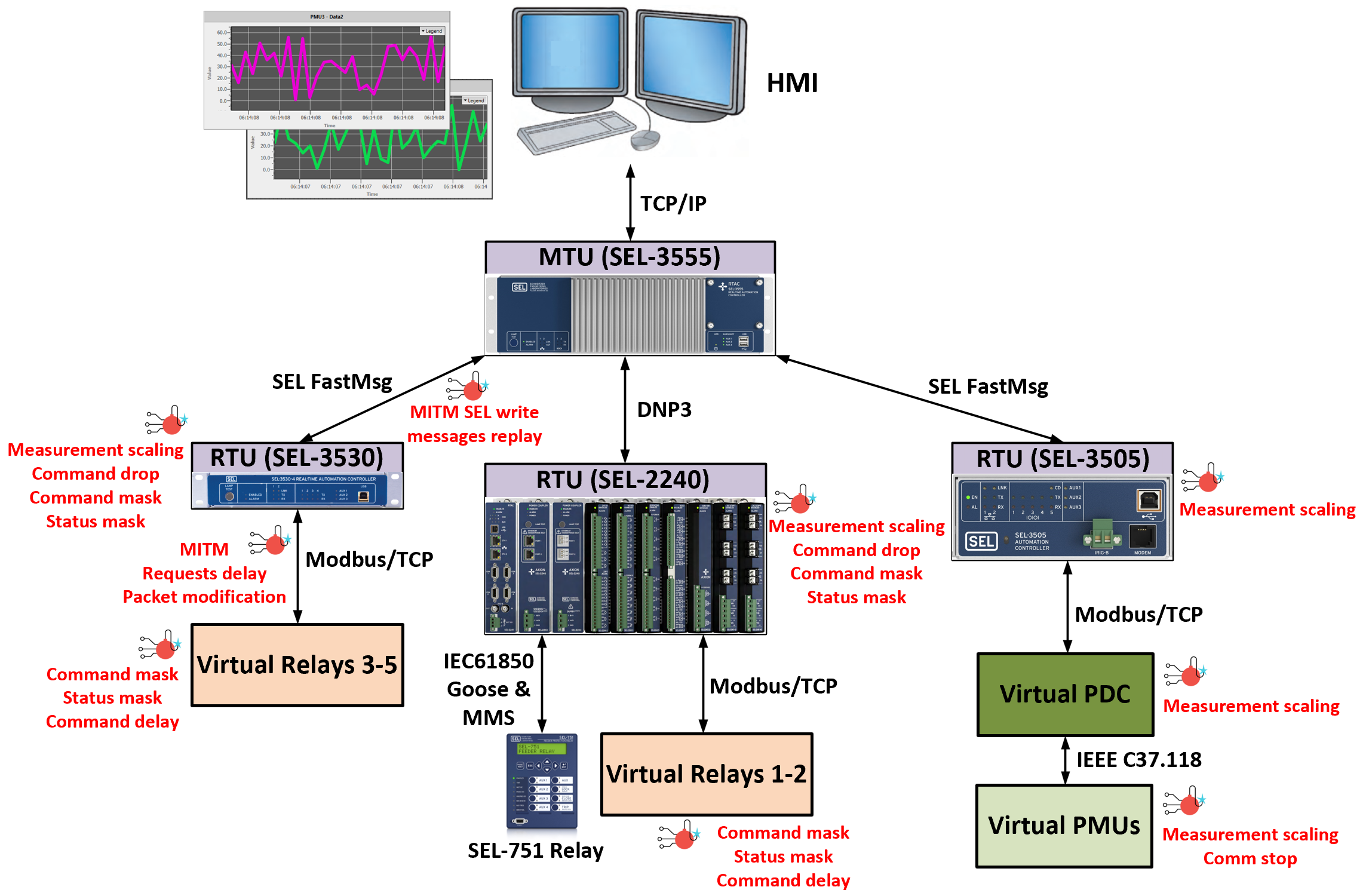}
  \caption{Physical and virtual nodes and roles in HIL testbed. SEL RTACs are used along with simulated Relays, PMUs, and PDC communicating with a variety of different protocols. Various attacks that were tested are shown in red between devices.}
  \label{fig:HIL_communication_setup_and_attacks}
\end{figure*}

As a sample power system scenario for experimental testing, we defined a simple 4-bus topology shown in \ref{fig:MATLAB_power_system_simulation}.
The substation considered for monitoring in TRAPS is the set of components on the right side of Bus 1 in the figure. Area 1 is considered a remote area connected via a 3-phase transmission line.
There are 5 relays in the substation. Two loads are being fed (Loads 1 and 2) with breakers in a 1.5 breaker scheme. There are 4 PMUs (one for each bus).
The relays, PMUs, and PDC are implemented as virtual nodes  (Figure~\ref{fig:HIL_communication_setup_and_attacks}) while the RTACs and HMI are physical devices. These physical and virtual nodes communicate using a variety of protocols including DNP3, Modbus/TCP, IEEE C37.118, IEC61850 Goose and MMS, and SEL Fast Msg, as shown in Figure~\ref{fig:HIL_communication_setup_and_attacks}.
To model the behavior of the overall CPS, a set of raw tags was defined as summarized in Table~\ref{tab:tags} in terms of which several match/pre/post/threshold conditions were listed
as illustrated in Table~\ref{tab:conditions}
based on the defined roles and communications of the physical and virtual roles in the system.
Each tag filtering rule and condition are related to a specific part of the CPS design shown in Figure~\ref{fig:HIL_communication_setup_and_attacks}. For example, the tag at Index 4 keeps track of the Modbus Write Single Register Requests transmitted from SEL-2240 to Relay 1 and 2, which are holding registers with Integer type.
As examples of conditions, observe that Condition 3 is a match requirement between values of tags 1 and 2 (faithful forwarding of PMU measurements by a PDC). On the other hand, Condition 8 is a post-condition between tags 12 and 4 (correct relaying of a HMI relay command by the RTAC).
Several more tags and conditions could be defined analogously to capture characteristics such as power flows in different parts of the system (as computed tags), time-averaged or low pass filtered signals calculated from power measurements, temporal patterns of tag observations, etc.


\begin{table*}
    \centering
    \begin{tabular}{|ccccc|}
        \hline
        \textbf{Index} & \textbf{Source} & \textbf{Destination} & \textbf{Protocol and Msg Type} & \textbf{Data type} \\
        \hline\hline
        \textbf{1} & PMU$_{1-4}$ & PDC & IEEE C37.118 Data Frame & Phasor, Analog and Digital data \\
        \textbf{2} & PDC & SEL-3505 & Modbus Read Holding Register Response & Holding Registers \\
        \textbf{3} & SEL-3505 & SEL-3555 & SEL Fast Msg Unsolicited Write & Float Registers \\
        \textbf{4} & SEL-2240 & Relay$_{1-2}$ & Modbus Write Single Register Request & Int Holding Registers \\
        \textbf{5} & SEL-2240 & Relay$_{1-2}$ & Modbus Read Holding Register Request & - \\
        \textbf{6} & Relay$_{1-2}$ & SEL-2240 & Modbus Write Single Register Response & Int Holding Registers \\
        \textbf{7} & Relay$_{1-2}$ & SEL-2240 & Modbus Read Holding Register Response & Int and Float Holding Registers \\
        \textbf{8} & SEL-3530 & Relay$_{3-5}$ & Modbus Write Single Register Request & Int Holding Registers \\
        \textbf{9} & SEL-3530 & Relay$_{3-5}$ & Modbus Read Holding Register Request & - \\
        \textbf{10} & Relay$_{3-5}$ & SEL-3530 & Modbus Write Single Register Response & Int Holding Registers \\
        \textbf{11} & Relay$_{3-5}$ & SEL-3530 & Modbus Read Holding Register Response & Int and Float Holding Registers \\
        \textbf{12} & SEL-3555 & SEL-2240 & DNP3 Operate Request & Group 41 all variations Int Analog Output \\
        \textbf{13} & SEL-3555 & SEL-2240 & DNP3 Read Request & Group 60 all variations Int Analog Input \\
        \textbf{14} & SEL-2240 & SEL-3555 & DNP3 Operate Response & Group 41 all variations Int Analog Output \\
        \textbf{15} & SEL-2240 & SEL-3555 & DNP3 Read Response & Group 30 all variations Int and Float Analog Inputs \\
        \textbf{16} & SEL-3555 & SEL-3530 & SEL Fast Msg Unsolicited Write & Int Registers \\
        \textbf{17} & SEL-3530 & SEL-3555 & SEL Fast Msg Unsolicited Write & Int and Float Registers \\
        \textbf{18} & Any Device & Relay$_{1-2}$ & Modbus Write Single Register Request & Int Holding Registers \\
        \textbf{19} & Any Device & Relay$_{3-5}$ & Modbus Write Single Register Request & Int Holding Registers \\
        \textbf{20} & SEL-2240 & SEL-751 & IEC61850 GOOSE Multicast & Float Analog Registers \\
        \textbf{21} & SEL-2240 & SEL-751 & IEC61850 MMS Report Unbuffered & Float Analog Registers \\
        \textbf{22} & SEL-751 & SEL-2240 & IEC61850 GOOSE Multicast & Float Analog Registers \\
        \textbf{23} & SEL-751 & SEL-2240 & IEC61850 MMS Report Unbuffered & Float Analog Registers \\
        \hline
    \end{tabular}
    \caption{Sample raw tags for extracting semantic observations from network traffic in the HIL simulation setup. Register address details  are omitted for brevity.}
\label{tab:tags}
\end{table*}

\begin{table*}
    \centering
    \begin{tabular}{|ccccc|}
        \hline
        \textbf{Index} & \textbf{Condition type} & \textbf{Tag 1 ID} & \textbf{Tag 2 ID} & \textbf{Threshold} \\
        \hline\hline
        \textbf{1} & Threshold condition on value & 1 & - & 1\% \\
        \textbf{2} & Threshold condition on time & 1 & - & 0.05s-0.15s \\
        \textbf{3} & Match condition & 1 & 2 & 1\% \\
        \textbf{4} & Match condition & 2 & 3 & 1\% \\
        \textbf{5} & Post-condition & 4,8 & 7,11 & 1.1s, 1\% \\
        \textbf{6} & Match condition & 7 & 15 & 1\% \\
        \textbf{7} & Post-condition & 12 & 15 & 6.0s, 1\% \\
        \textbf{8} & Post-condition & 12 & 4 & 0.3s, 1\% \\
        \textbf{9} & Match condition & 11 & 17 & 1\% \\
        \textbf{10} & Post-condition & 16 & 17 & 1.1s, 1\% \\
        \textbf{11} & Post-condition & 16 & 8 & 0.3s, 1\% \\
        \textbf{12} & Threshold condition on time & 9 & - & 0.98s-1.02s \\
        \textbf{13} & Threshold condition on time & 17 & - & 0.95s-1.05s \\
        \textbf{14} & Pre-condition & 18 & 12 & 0.3s, 1\% \\
        \textbf{15} & Pre-condition & 19 & 16 & 0.3s, 1\% \\
        \textbf{16} & Match condition & 22,23 & 15 & 1\% \\
        \textbf{17} & Post-condition & 12 & 20,21 & 1.0s, 1\% \\
        \hline
    \end{tabular}
    \caption{Sample conditions to model the CPS temporal behavior.
      The defined thresholds are based on the nominal communication characteristics and power system dynamics.
      For the threshold conditions on value and match conditions, the defined threshold shows the allowed deviation from nominal values.
      For pre-/post-conditions, the defined threshold shows the allowed time window and deviation from nominal values, respectively.
    }
\label{tab:conditions}
\end{table*}

\subsection{Evaluation of Attack Detection} \label{sec:Attacks}
To test the attack detection performance of the proposed framework, a wide range of
attacks were implemented in our HIL setup as shown in red in Figure~\ref{fig:HIL_communication_setup_and_attacks}.
Our experiments mainly focus on adversarial manipulations of the grid's devices' logic and behavior, which are among the most potent and stealthy attacks that could be done via supply chain, firmware tampering, and advanced persistent threat (APT) lateral movement attacks. 
The considered attacks summarized in Table~\ref{tab:scenarios} cover a vast range of real-world attacks. The attacks on the PMUs, PDC, RTACs, and relays are simulated as malicious firmware delivered via supply chain attacks. On the other hand, the MITM attacks on Modbus/TCP and the SEL Fast Msg protocols are performed by injecting an external device in the path of communication between devices. An ODROID-XU4 was used as the MITM device and an MITM interception was implemented using the Python Scapy library and Scapy Modbus package (for intercepting and modifying Modbus requests and response packets). MITM attacks could alternatively be performed using ARP spoofing, which results in ARP cache poisoning on the victim devices and enables the attacker to intercept communications. In our MITM implementation, an intruder device is physically placed in the communication path, therefore removing need for ARP spoofing and making the attack even more effective and stealthy.

\begin{table}
    \centering
    \begin{tabular}{|cp{2.8in}|}
        \hline
        \textbf{Index} & \textbf{Scenario} \\
        \hline\hline
        \textbf{1} & FDI attack in virtual PMUs through measurement scaling\\
    \textbf{2} & DoS attack in virtual PMUs through communication disruption\\
    \textbf{3} & FDI attack on the PDC in the form of measurements scaling \\
    \textbf{4} & FDI attack in SEL-3505 using measurements scaling \\
    \textbf{5} & FDI and FCI attacks in Virtual relays through commands and status masking \\
    \textbf{6} & DoS attack in virtual relays in the form of command delaying \\
    \textbf{7} & FDI attack on SEL-2240 in the form of measurement scaling \\
    \textbf{8} & FDI and FCI attacks on SEL-2240 through status and command masking \\
    \textbf{9} & DoS attack on SEL-2240 by dropping commands \\
    \textbf{10} & FDI attack on SEL-3530 through measurement scaling \\
    \textbf{11} & FDI and FCI attacks on SEL-3530 through status and command masking \\
    \textbf{12} & DoS attack on SEL-3530 by dropping commands\\
    \textbf{13} & MITM attack on Modbus/TCP protocol between the virtual relays and SEL-3530, which includes DoS in the form of Read Holding Registers request delaying and FDI through packet modification \\
    \textbf{14} & MITM attack on SEL Fast Msg protocol between the SEL-3530 and SEL-3555, which includes replay attack on SEL Unsolicited Write messages\\
    \textbf{15} & FCI attack on Relay3-5 by sending false commands from SEL-3530 (or any other intruder devices in the network) to open/close the relay without the request of HMI\\
        \hline
    \end{tabular}
    \caption{Various attack scenarios considered in HIL testing.
    }
\label{tab:scenarios}
\end{table}

\begin{figure*}[!t]
  \centering
    \begin{subfigure}{0.49\textwidth}
        \includegraphics[width=\textwidth]{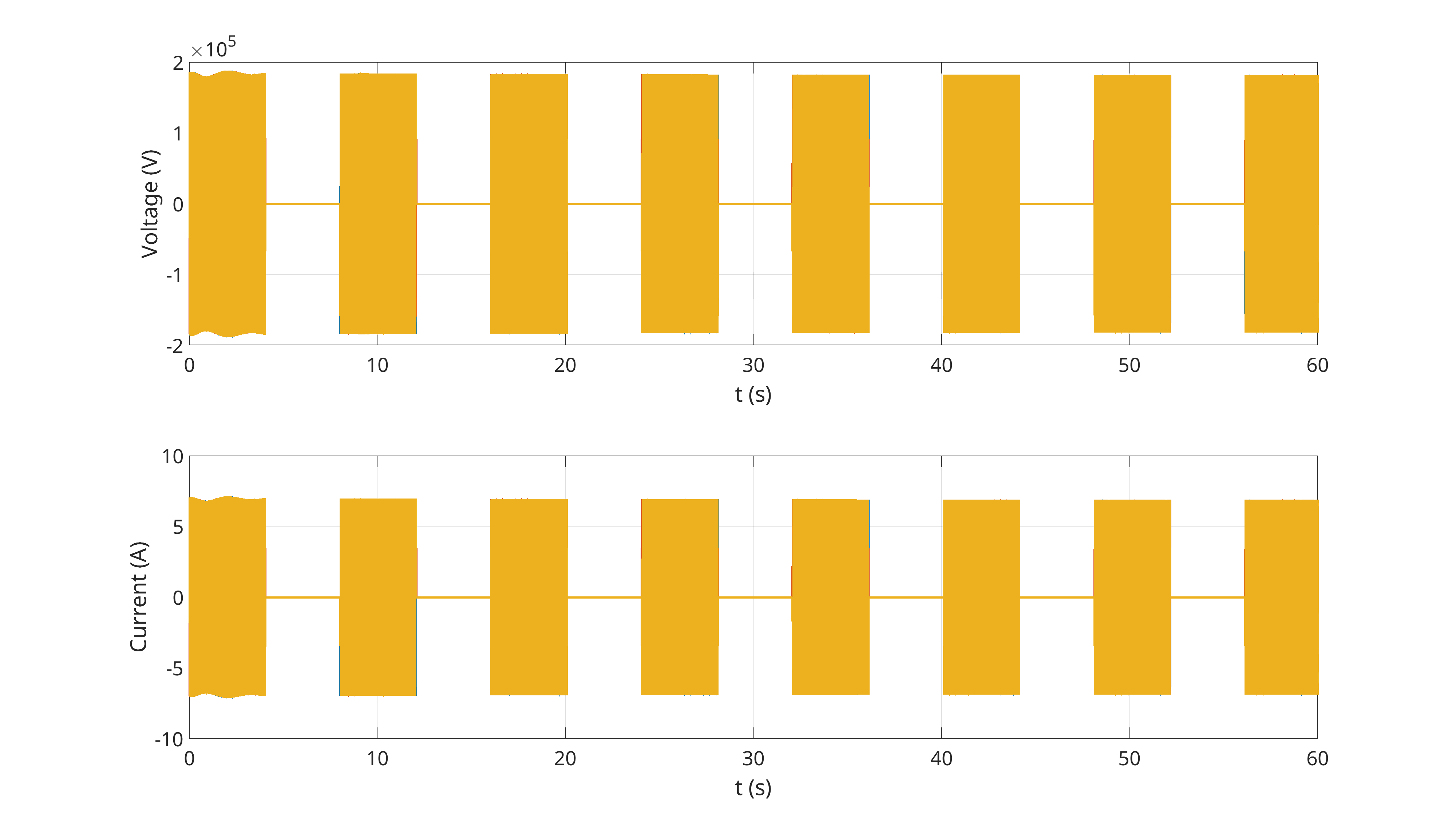}
    \end{subfigure}
    \begin{subfigure}{0.49\textwidth}
        \includegraphics[width=\textwidth]{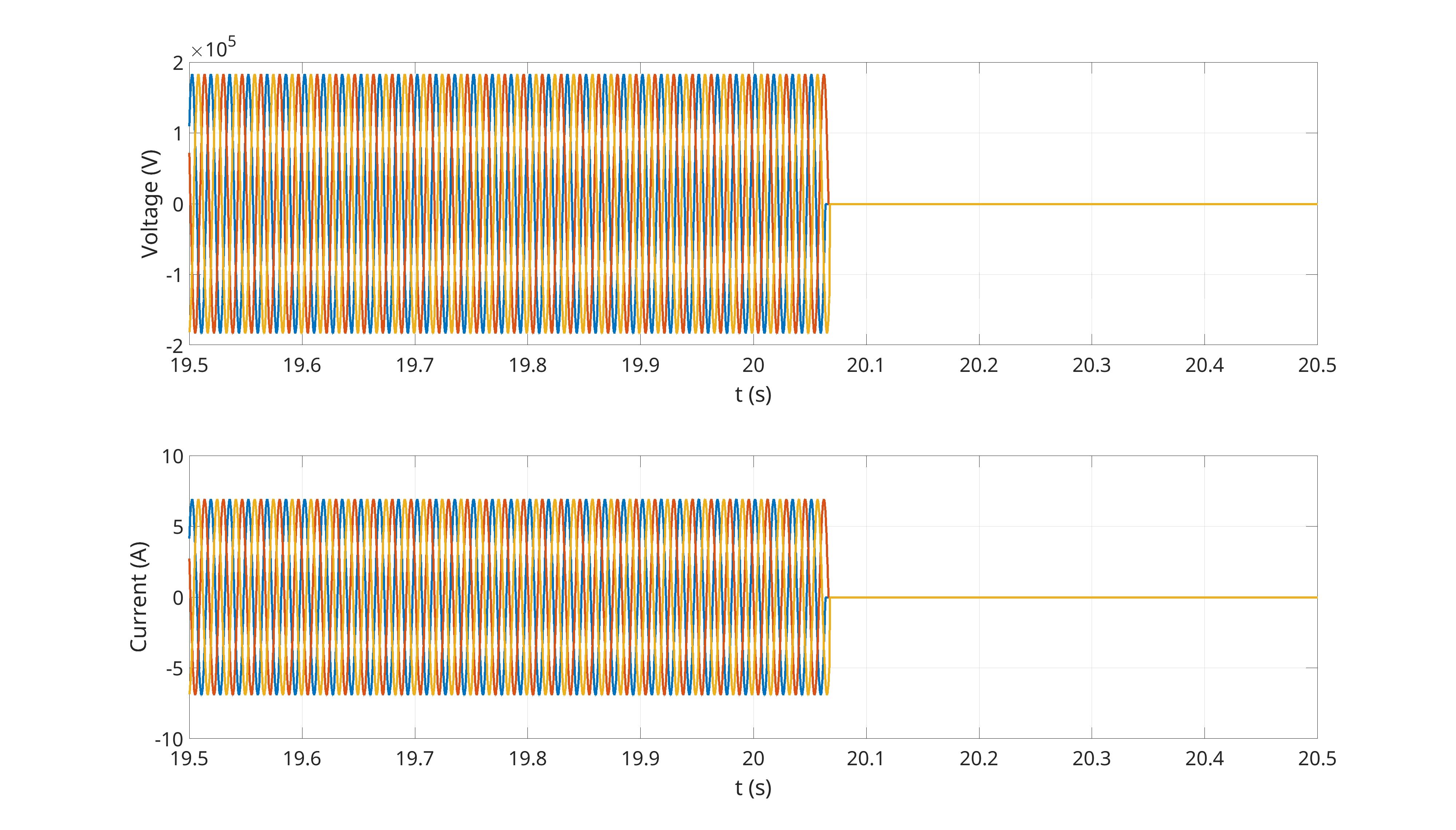}
    \end{subfigure}
  \caption{Voltage (phase to ground) and current plots for Load 1 in Figure \ref{fig:MATLAB_power_system_simulation} (as measured at Bus 3) when the attacker sends commands to Relays 3 and 4 in Figure \ref{fig:HIL_communication_setup_and_attacks} to cycle them on and off with a time interval of 4 seconds between commands. This has the effect of cycling power to Load 1. Voltage and current plots for other buses are omitted for brevity. The right-side figure is a zoomed-in view of the plot in the left-side figure.}
  \label{fig:Substation_Relay1_2_attack_plots}
\end{figure*}

We captured 5 minutes of network communication traffic from the HIL simulator for the normal mode and when triggering each of the attacks listed in Table~\ref{tab:scenarios}. We fed each of these captured traffic data to our proposed TRAPS system and observed that all the considered attack scenarios are reliably detected after transitioning from the normal mode to the attack mode since at least one of the conditions in Table~\ref{tab:conditions} is violated under each scenario.
For example, attack scenario 9 (Command dropping attack on SEL-2240) results in violation of Post-condition 8 between Tags 12 and 4, which states that the value of Tag 12 (DNP3 Operate Requests from SEL-3555 to SEL-2240) should match with Tag 4 (Modbus Write Single Register Request from SEL-2240 to Relay 1-2) within a short time window.
As another example, attack scenario 15, which is an FCI attack to Relays 3 and 4 in the form of sending false open/close commands to relays from SEL-3530 (or any other intruder devices) is depicted in Figure~\ref{fig:Substation_Relay1_2_attack_plots}.
This attack is detected using Pre-condition 15 since it requires that if a Modbus write single Register Request is being sent to Relays 3-5 in Figure~\ref{fig:HIL_communication_setup_and_attacks}, then within a short time window before that, a corresponding command should have been sent from SEL-3555 to SEL-3530 (i.e., an authorized command by the HMI) requesting the relays to be opened or closed. This condition is not satisfied when the attacker sends false messages, leading to detection of an anomaly.

\begin{figure}[!t]
  \centering
  \includegraphics[width=0.7\linewidth]{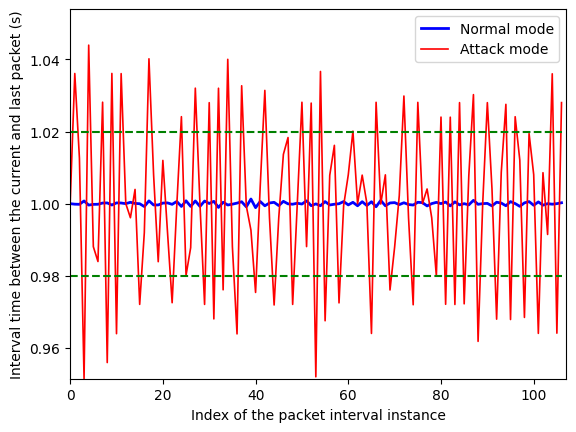}
  \caption{Plot of interval times between subsequent Modbus Read Holding Register Requests from SEL-3530 to Relay3 in Figure \ref{fig:HIL_communication_setup_and_attacks} in normal mode and MITM attack mode. The interval times have a relatively large variance in the attack case, enabling detection of the anomaly/attack using the defined threshold conditions. The horizontal dashed lines show the thresholds defined on the value of time intervals in the conditions of Table~\ref{tab:conditions}.}
  \label{fig:MITM_delay_attack_plots_1}
\end{figure}

As another illustration of anomaly detection by the proposed framework, the timing plots under attack scenario 13, which is an MITM attack between SEL-3530 and Relays 3-5 are shown in Figure~\ref{fig:MITM_delay_attack_plots_1}. The attacker randomly adds delays in the communication of Modbus Read Holding Registers Request messages sent from SEL-3530 to Relays 3-5 in Figure \ref{fig:HIL_communication_setup_and_attacks}.
As seen in Figure~\ref{fig:MITM_delay_attack_plots_1}, the time intervals between subsequent Modbus request packets under the MITM attack have an anomalous temporal pattern triggering raising of an alert based on condition 12 in Table~\ref{tab:conditions}.

\begin{figure}[!t]
  \centering
  \includegraphics[width=0.7\linewidth]{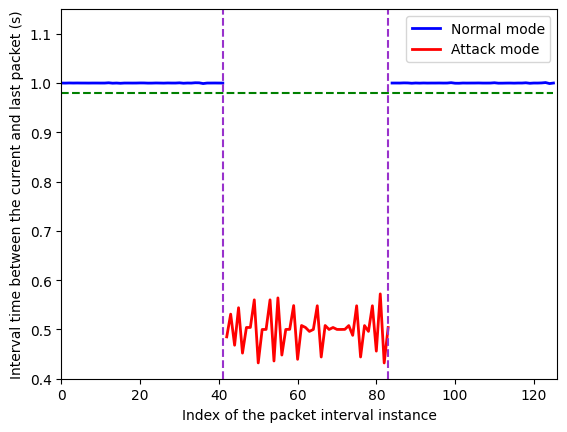}
  \caption{Plot of interval times between subsequent SEL Fast Msg Unsolicited Write messages from SEL-3530 to SEL-3555 in Figure~\ref{fig:HIL_communication_setup_and_attacks} in normal mode and MITM replay attack mode. In the case of attack mode, the delay value drops lower than the defined threshold in the conditions of Table~\ref{tab:conditions},  resulting in detection of an anomaly.}
  \label{fig:MITM_replay_attack_plots_1}
\end{figure}

Another example attack is illustrated in Figure~\ref{fig:MITM_replay_attack_plots_1}, which shows the timing plots under attack scenario 14 (MITM replay attack between SEL-3530 and SEL-3555). In this case, the attacker retransmits some SEL Fast Msg packets From SEL-3530 to SEL-3555. The figure shows time intervals between subsequent SEL Unsolicited Write messages during normal and attack modes.
These timings are detected as anomalous under the attack mode based on Condition 13 of Table~\ref{tab:conditions}.

While the above examples illustrate some types of attacks detected by our proposed TRAPS framework, an important point to note is that TRAPS is not designed specifically to detect these (or any other) particular attacks. Rather, TRAPS is intended as a flexible behavioral modeling and integrity verification framework in which the semantic temporal characteristics of the CPS based on the design logic, physics, communication configuration, etc., can be defined to any desired level of detail and thereafter {\em automatically} monitored in real-time by TRAPS and any deviations flagged as anomalies.
A key part of enabling such a flexible framework is scalability and computational tractability. To evaluate these aspects, the timing characteristics of the current TRAPS implementation were measured as shown in Table~\ref{tab:timing_perf}. The average of the total processing time for the packets in the mentioned setup in a 5-minute capture is about 60 $\mu s$, which means that the proposed IDS is capable of processing around 16,600 packets per second. Considering the average packet length of 143 bytes in our captures, this corresponds to about a throughput of 19 Mbps (considering TCP handshakes, etc., the effective throughput is actually around 25 Mbps). Note that these timing numbers are with our current implementation which is purely in Python (for ease of prototyping) and considerable speed-ups can be naturally expected by switching to a faster implementation language (e.g., C++), further enhancing scalability of the proposed framework.  

\begin{table*}
    \centering
    \begin{tabular}{|cc|}
        \hline
        \textbf{Processing Item} & \textbf{Timing Performance} \\
        \hline\hline
        Average total processing time & 60.0 $\mu s$ \\
        Average processing time for raw tags per tag & 1.5 $\mu s$ \\
        Average processing time for raw tags per matched tag & 58.0 $\mu s$ \\
        Average processing time for computed tags per tag & 0.87 $\mu s$ \\
        Average processing time for computed tags per matched tag & 21.0 $\mu s$ \\
        Average total processing time for conditions per tag & 12.3 $\mu s$ \\
        Average processing time for match-conditions per tag per condition & 0.46 $\mu s$ \\
        Average processing time for pre-conditions per tag per condition & 0.42 $\mu s$ \\
        Average processing time for post-conditions per tag per condition & 0.62 $\mu s$ \\
        Average processing time for threshold conditions per tag per condition & 1.0 $\mu s$ \\
        Average latency to detect anomalies per packet with anomaly & 2.5 $ms$ \\
        \hline
    \end{tabular}
    \caption{Timing performance of the proposed anomaly monitoring system for each packet.}
\label{tab:timing_perf}
\end{table*}

\section{Conclusion} \label{sec:Conclusion}
A novel anomaly monitoring and integrity verification framework for CPS based on dynamic
behavioral analysis was developed and experimentally demonstrated on a HIL
testbed emulating a smart grid SCADA system. The efficacy of the framework was shown under a wide range of attack
scenarios that span the commonly seen categories such as FDI, FCI, MITM, and DoS
attacks. The proposed framework is designed to be scalable and computationally
lightweight to facilitate applicability to real-time monitoring of CPS as seen
from the timing/throughput characteristics of the prototype implementation
(which can be further optimized through, for example, multithreading and re-implementation in a
compiled language such as C++ instead of Python).
Future work will address further integration of time-series processing
techniques and learning-based methods into the framework (both for learning of
the tag definitions and conditions to reduce operator workload during
configuration and for learning of baseline operating characteristics relative to
which anomalies are detected for change detection).

\section*{Acknowledgments} \label{sec:Acknowledgments}
The authors would like to thank collaborators from Narf (Prashant Anantharaman, Michael Locasto, and others) and SRI (Nick Boorman, Ulf Lindqvist) for their work on parsers for several network communication protocols used in experiments in this study as well as many helpful discussions.

\bibliographystyle{IEEEtran}
\bibliography{references}
\end{document}